\documentclass[preprint,review,12pt]{elsarticle}

\usepackage{lineno,hyperref}
\usepackage{amsmath}
\usepackage{prettyref}
\usepackage{subfigure}
\usepackage{float}
\usepackage{booktabs}

\hypersetup{hyperindex=false, colorlinks=true, linkcolor=blue, citecolor=blue, urlcolor=blue}
\newrefformat{fig}{Fig.~\ref{#1}}
\newrefformat{tbl}{Table~\ref{#1}}
\newrefformat{sec}{Section~\ref{#1}}
\newrefformat{eq}{Eq.~\ref{#1}}
\newrefformat{alg}{Algorithm~\ref{#1}}

\makeatletter
\def\ps@pprintTitle{%
   \let\@oddhead\@empty
   \let\@evenhead\@empty
   \def\@oddfoot{\reset@font\hfil\thepage\hfil}
   \let\@evenfoot\@oddfoot
}
\makeatother

\biboptions{authoryear}

\begin{document}

\begin{frontmatter}

\title{Outcrop fracture characterization on suppositional planes cutting through digital outcrop models (DOMs)\tnoteref{t1}}
\tnotetext[t1]{Code available from: https://github.com/EricAlex/structrock.}

\author[zju]{Xin Wang}
\ead{ericrussell@zju.edu.cn}

\author[zju]{Lejun Zou}

\author[zju]{Yupeng Ren}

\author[zju]{Yi Qin}

\author[zju]{Zhonghao Guo}

\author[zju]{Xiaohua Shen}
\ead{shenxh@zju.edu.cn}

\address[zju]{School of Earth Sciences, Zhejiang University Yuquan Campus, 38 Zheda Road, Hangzhou 310027, China}


\begin{abstract}

Conventional fracture data collection methods are usually implemented on planar surfaces or assuming they are planar; these methods may introduce sampling errors on uneven outcrop surfaces. Consequently, data collected on limited types of outcrop surfaces (mainly bedding surfaces) may not be a sufficient representation of fracture network characteristic in outcrops. Recent development of techniques that obtain DOMs from outcrops and extract the full extent of individual fractures offers the opportunity to address the problem of performing the conventional sampling methods on uneven outcrop surfaces. In this study, we propose a new method that performs outcrop fracture characterization on suppositional planes cutting through DOMs. The suppositional plane is the best fit plane of the outcrop surface, and the fracture trace map is extracted on the suppositional plane so that the fracture network can be further characterized. The amount of sampling errors introduced by the conventional methods and avoided by the new method on 16 uneven outcrop surfaces with different roughnesses are estimated. The results show that the conventional sampling methods don't apply to outcrops other than bedding surfaces or outcrops whose roughness $>$ 0.04 m, and that the proposed method can greatly extend the types of outcrop surfaces for outcrop fracture characterization with the suppositional plane cutting through DOMs.

\end{abstract}


\end{frontmatter}


\section{Introduction}

The characterization of fractures in outcrops is important in many areas of geology (e.g., structural and geomechanical analysis, reservoir characterization, and engineering rock mass classification) as the abundance and arrangement of the fractures may control many of the physical properties of rocks, such as stiffness, strength, porosity and permeability \citep{Adler99}. Although fracture systems occupy three dimensional volumes, they are most often studied in one-dimensional transects or two-dimensional maps and cross section \citep{Olariu08}.

Four main conventional fracture data collection methods are widely used and reported in the literature: the linear scanline method \citep{Priest81,Priest93}, areal sampling \citep{Wu95}, rectangular window sampling \citep{Pahl81,Priest93} and the circular scanline method \citep{Mauldon01,Rohrbaugh02}. Those methods are usually implemented on planar surfaces (such as bedding surfaces) or assuming they are planar. Attributes of individual fractures such as orientation, size, morphology, etc., the statistical distribution of the attributes and the topology of the fracture network (as proposed by \cite{Sanderson15}) are measured and characterized on those surfaces.

However, those existing methods don't apply to outcrops that have uneven surfaces, as is often the case when they are not bedding surfaces, because of sampling errors that may be introduced by sampling on uneven surfaces using methods that were designed to perform on planar surfaces. In addition, in regions of three-dimensional heterogeneous fracture network systems, data collected on limited types of outcrop surfaces (mainly bedding surfaces) may not be a sufficient representation of fracture network characteristic in outcrops.

The development of remote sensors (e.g., LiDAR-based scanners), their availability as research equipment and recent development of methods that process the data acquired by the equipment from outcrop may provide an opportunity of developing new methods to address the problems of the conventional sampling methods as mentioned above. The terrestrial laser scanner (TLS) have proven very useful for acquisition of high-quality, high-resolution, three-dimensional (3D) terrain data from outcrops \citep[e.g.,][]{Xu00,Mccaffrey05,Pollyea11,Mah13}. The digital outcrop models (DOMs) \citep{Bellian05} and many attributes of individual fractures such as orientation, size, morphology, etc., can be measured or extracted from the scanned terrain data from outcrops. Many semi-automatic or automatic methods for the extraction of fracture attributes have been developed in the last 10 years. Recently, \cite{Wang17} proposed a method that can extract the full extent of every individual fracture on the scanned outcrop surface.

In this study, based on the DOMs and the full extent of every individual fracture, we developed a new method to address the problems of the conventional methods as mentioned above. In this method the outcrop fracture network is characterized on suppositional planes cutting through DOMs. The types of sampling errors introduced by the conventional methods and avoided by the proposed method are discussed. The amount of sampling errors avoided by our method are estimated on 16 outcrops with different roughnesses to find conditions where the conventional methods do not apply and the proposed method will be performed, and to show the advantages of our method, thereby to demonstrate the high potential utility of the proposed method.

\section{Methodology}

\subsection{DOMs and the full extent of individual fractures}

DOMs are triangulated irregular networks (TIN), or point clouds, collections of 3D points $\boldsymbol{p} = \{x, y, z\}$, generated from the terrain data of outcrops acquired by TLSs, and our method uses mainly the ``point clouds'' form of the DOMs. From the point clouds, the method proposed by \cite{Wang17} extracts the full extent of individual fractures as patches of points that form the fracture surfaces. Attributes of those fractures thus can be extracted and further characterization of the fracture network can be performed in the following manner.

\subsection{The suppositional plane cutting through DOMs}

For uneven outcrop surfaces to which the conventional sampling methods don't apply, having their DOMs and the full extent of individual fractures, we introduce the suppositional plane cutting through DOMs. The suppositional plane serves as a planar surface on which the conventional sampling methods can be performed.

Informations of outcrop fractures on the suppositional plane should be found or estimated to perform further fracture characterization. The suppositional plane is at the position that best fit the outcrop surface, i.e., as close to the outcrop surface as possible, for the convenience of finding or estimating fracture informations on the suppositional plane as accurately as possible, since the full extent of individual fractures are all on the outcrop surface.

We refer to the point clouds form of DOMs as $P$, a collection of 3D points $\boldsymbol{p}_i = \{x_i, y_i, z_i\} \in P; i \in \{1, 2, 3, ... N\}$. The method we used for finding the suppositional plane that best fit the outcrop surface (represented by $P$) is based on least-squares plane fitting with $P$, as proposed by \cite{Berkmann94}. $P$'s covariance matrix $C = \frac{1}{N}\sum_{i=1}^N (\boldsymbol{p}_i - \bar{\boldsymbol{p}}) \cdot (\boldsymbol{p}_i - \bar{\boldsymbol{p}})^\mathrm{T}$, where $\boldsymbol{p}_i \in P$ and $\bar{\boldsymbol{p}} = \frac{1}{N} \cdot \sum_{i=1}^N \boldsymbol{p}_i$, let $\lambda_0$, $\lambda_1$, and $\lambda_2$ be the eigenvalues of $C$ that satisfy $0 \leq \lambda_0 \leq \lambda_1 \leq \lambda_2$ and if $\boldsymbol{v}_0$ is the corresponding eigenvector of $\lambda_0$, then we have the suppositional plane $S$: $\boldsymbol{v}_0 \cdot (\boldsymbol{x}-\bar{\boldsymbol{p}}) = 0$, where $\boldsymbol{x} = \{x, y, z\}$.

\subsection{The extraction of fracture trace map on the suppositional plane}

In the characterization of fracture networks on the suppositional plane $S$, such as the topology of the fracture network, the fracture trace map on $S$ is needed. First, the 2D boundary polygon (e.g. 2D convex hull) of the outcrop on $S$ within which the fracture trace map will be extracted needs to be defined. The 2D boundary polygon is basically the boundary of the outcrop point cloud $P$'s ``shadow'' on $S$, and let $Q$ be the projection of $P$ on $S$, the 2D convex hull of $Q$ needs to be found. Qhull \citep{Barber96}, an open source library, is used to compute the 2D convex hull.

We already know that $\bar{\boldsymbol{p}}$ is the center of $P$ and that $\boldsymbol{v}_0$ is the normal vector of $S$, and if $\boldsymbol{q}_i \in Q$ on $S$ is the projected point of $\boldsymbol{p}_i \in P$, we have $\boldsymbol{q}_i = \boldsymbol{p}_i - ((\boldsymbol{p}_i - \bar{\boldsymbol{p}}) \cdot \boldsymbol{v}_0)\boldsymbol{v}_0$. The method used to compute $Q$'s convex hull $C$ is described in \cite{Barber96}.

Fracture traces are intersections between the fracture faces and the sampling surface. Most extracted fracture faces already have intersections with $S$, but some of the fractures may have to extend toward $S$ to intersect with it (can be found on very uneven outcrop surfaces). The full extent of fracture faces extracted from the point cloud are only the exposed part of the whole fracture surfaces. We judged that the intersections between the extracted fracture faces and $S$ may not be a sufficient representation of the fracture length. The same rule of extending fracture faces to find their fracture traces on $S$ is applied to all the fracture faces no matter they already have intersections with $S$ or not. The fracture face only extends in the direction of reaching $S$, and the extension will result in a fracture trace with its length equals to the width of the fracture face with respect to the extending direction. With all the fracture traces derived from the same extension rule and that are not outside the outcrop convex hull $C$, we have the fracture trace map on $P$.

Let $P^m$ be a patch of points that form the full extent of a fracture face, $S_i$ be the best fit plane in $P^m$ found using method described above, $\boldsymbol{n}_i$ be the normal vector of $S_i$, $Q^m$ be the set of points on $S_i$ projected from $P^m$ using method described above, and $\boldsymbol{v}_0 \times \boldsymbol{n}_i$ be the y-axis on $S_i$, then for points in $Q^m$, we have $\boldsymbol{q}_\mathrm{max}$ with the maximum $y$ value $y_\mathrm{max}$ and $\boldsymbol{q}_\mathrm{min}$ with the minimum $y$ value $y_\mathrm{min}$. The width of the fracture face is $y_\mathrm{max} - y_\mathrm{min}$.

The extending direction is $\boldsymbol{n}_i \times \boldsymbol{v}_0 \times \boldsymbol{n}_i$. If the line with the direction $\boldsymbol{n}_i \times \boldsymbol{v}_0 \times \boldsymbol{n}_i$ goes through $\boldsymbol{q}_\mathrm{max}$, and crosses $S$ at $\boldsymbol{j}_\mathrm{max}$, and the line with the direction $\boldsymbol{n}_i \times \boldsymbol{v}_0 \times \boldsymbol{n}_i$ goes through $\boldsymbol{q}_\mathrm{min}$, and crosses $S$ at $\boldsymbol{j}_\mathrm{min}$, we have the fracture trace $\boldsymbol{j}_\mathrm{max}\boldsymbol{j}_\mathrm{min}$ if it's not outside the outcrop convex hull $C$.

\subsection{The workflow}

\begin{figure}[H]
  \centering
  \includegraphics[width=0.98\textwidth]{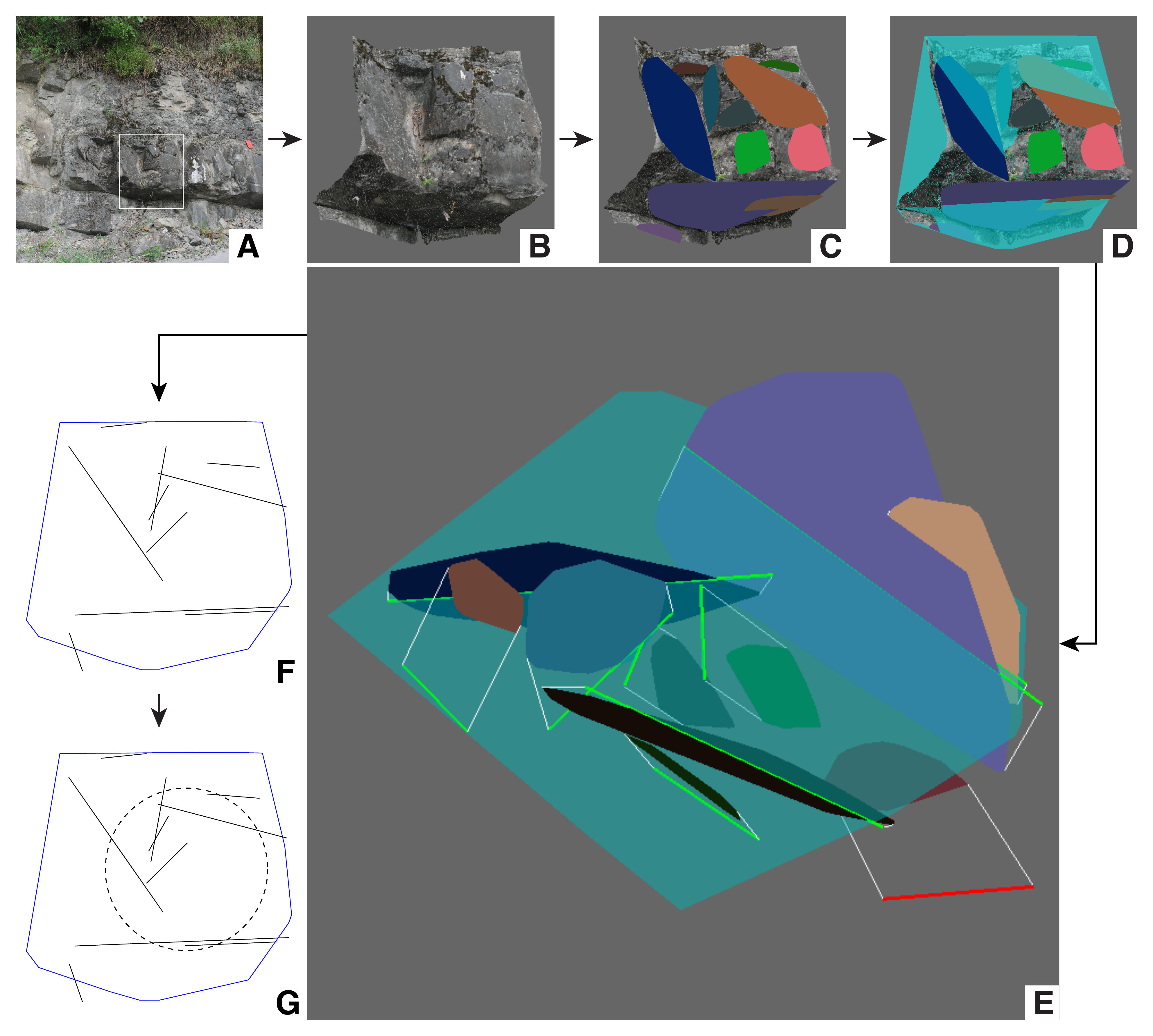}
  \caption{Workflow of the proposed method. A: Picture of the outcrop, the white rectangle highlights the example region. B: Digital outcrop models (DOMs) of the example region. C: The full extent of individual fractures: different fracture faces are indicated by different colors. D: The suppositional plane (the part inside the boundary polygon of the example region) cutting through DOMs and the fracture faces. E: The extraction of fracture trace map on the suppositional plane. The white lines demonstrate the extension of fracture faces toward the suppositional plane. The fracture traces are obtained from the same extension rule. Red lines indicate fracture traces outside the boundary polygon of the outcrop, otherwise fracture traces are indicated by green lines. F: Constructed fracture trace map on the suppositional plane. G: Performing fracture characterization (such as the dashed circular scanline) on the suppositional plane.}
  \label{fig:workflow}
\end{figure}

It may be easier to understand the workflow of the proposed method through the graphical demonstration of a very simple example, as shown in \prettyref{fig:workflow}. For a small example region of the outcrop shown in \prettyref{fig:workflow}A, we have its DOMs and the full extent of individual fractures extracted from the DOMs (in the point cloud form) are shown in \prettyref{fig:workflow}C: different fracture faces are indicated by different colors. The suppositional plane cutting through DOMs and the fracture faces is found and \prettyref{fig:workflow}D shows the part of the suppositional plane that is inside the boundary polygon of the example region. Then the fracture trace map on the suppositional plane is obtained by extending fracture faces with the same extension rule. \prettyref{fig:workflow}E shows the extension and the obtained fracture trace map: red lines indicate fracture traces outside the boundary polygon of the outcrop, otherwise fracture traces are indicated by green lines. Finally, we can perform fracture characterization on the suppositional plane, as shown in \prettyref{fig:workflow}G.

\section{Results and Discussion}

Although the true shapes of the fracture faces in the outcrop may be different from that of the extended fracture faces in the procedure of extracting the fracture trace map on the suppositional plane, it is almost impossible to find all the unique ways each fracture face extends in 3D space, and as the suppositional plane is as close to the outcrop surface as possible, we judged it's simple and reasonable to assume that the fracture length doesn't change too much a very small distance away. Our method estimates the position of the fracture trace as accurately as possible by extending fracture faces toward the suppositional plane and finding their intersections, and the accuracy of fracture trace position is very important in many aspects of fracture network characterization, such as the topology of the fracture network.

As described above, the proposed method applies to uneven outcrop surfaces on which the conventional sampling methods may introduce sampling errors. In the error analysis, the types of sampling errors introduced by the conventional methods and avoided by our method are analyzed in \prettyref{sec:types_of_sampling_errors_conventional_methods}. The amount of sampling errors avoided by our method are estimated in \prettyref{sec:the_estimation_of_sampling_errors} on outcrops with different roughnesses to find conditions where the conventional methods do not apply and the proposed method will be performed, and to show the advantages of the proposed method.

\subsection{The types of sampling errors introduced by the conventional methods}
\label{sec:types_of_sampling_errors_conventional_methods}

\begin{figure}[H]
  \centering
  \includegraphics[width=0.98\textwidth]{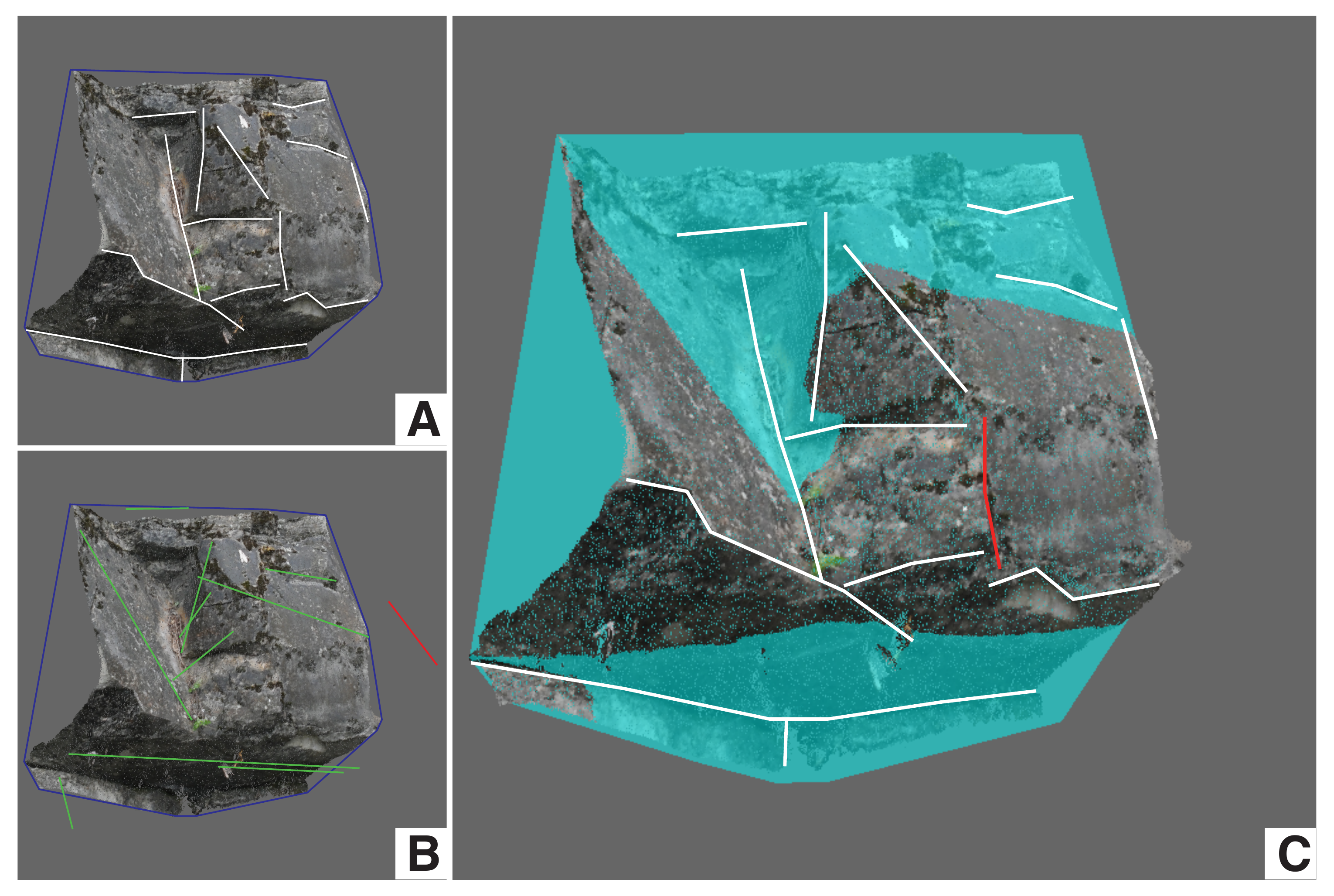}
  \caption{The fracture trace maps obtained from the conventional sampling methods and the proposed method, and the analysis of sampling errors introduced by the conventional methods. A: White lines are fracture traces obtained from the conventional sampling methods assuming the outcrop surface is planar. B: The fracture trace map obtained from our method. Red lines indicate fracture traces outside the boundary polygon of the example outcrop, otherwise fracture traces are indicated by green lines. C: The analysis of sampling errors introduced by the conventional sampling methods. The position and orientation of fracture traces on the sampling plane cutting thought the outcrop surface are inaccurately sampled. The red fracture trace is so displaced that its true position on the sampling plane is outside the boundary polygon of the outcrop.}
  \label{fig:error_analysis}
\end{figure}

In the same example region of the outcrop shown in \prettyref{fig:workflow}A, the conventional sampling methods of extracting the fracture trace map assuming the outcrop surface is planar and the proposed method are performed, and the results are shown in \prettyref{fig:error_analysis}A and B, respectively. As stated above, our method can accurately estimate a fracture trace's position and orientation. The assumption that the fracture length doesn't change too much a very small distance away may introduce some inaccuracy in the fracture length. But we judged the assumption and the estimated fracture length are statistically correct considering that the true fracture length can both be a little bit longer or shorter than the estimated fracture length. Although it's almost certain that our method estimates the fracture length more accurately and robustly than the conventional methods given the informations of the outcrop surface and the fracture faces, the advantages of our method over the conventional methods on the estimation of fracture trace's position and orientation are more obvious.

\prettyref{fig:error_analysis}C shows the fracture trace map extracted with the conventional sampling methods and the sampling plane cutting thought the outcrop surface to analyze the sampling errors. Three types of sampling errors are used to describe the sampling errors introduced by the conventional sampling methods: the displacement error measures the distance from incorrectly sampled fracture traces to correctly sampled fracture traces while the orientation error measures the difference between their orientations. The over sampling error measures the amount of fracture traces that should fall outside the boundary polygon of the outcrop and shouldn't be sampled, such as the red fracture trace in \prettyref{fig:error_analysis}C. All of the three types of sampling errors above can be avoided by the proposed method through the accurate estimation of fracture trace's position and orientation.

\subsection{The estimation of sampling errors avoided by the proposed method}
\label{sec:the_estimation_of_sampling_errors}

Although it's convenient to visualize and analyze the types of sampling errors introduced by the conventional sampling methods, extracting fracture trace map using the conventional sampling methods and comparing it with the sampling plane cutting thought the outcrop surface isn't a good way to estimate the amount of sampling errors. The variation in the fracture trace map introduced by the operator of the conventional sampling methods may cause unstable estimation of sampling errors. In addition, performing the conventional sampling methods to extract fracture trace maps can be very time-consuming and laborious, as many outcrops with different roughnesses need to be studied.

The orientation of fracture traces sampled using the conventional sampling methods depends on the complex geometry of the fracture face, the spatial relationship between neighboring fracture faces and even the operator. It's very difficult to estimate the orientation error by estimating the orientation of fracture traces the operator gets.

\begin{figure}[H]
  \centering
  \includegraphics[width=0.98\textwidth]{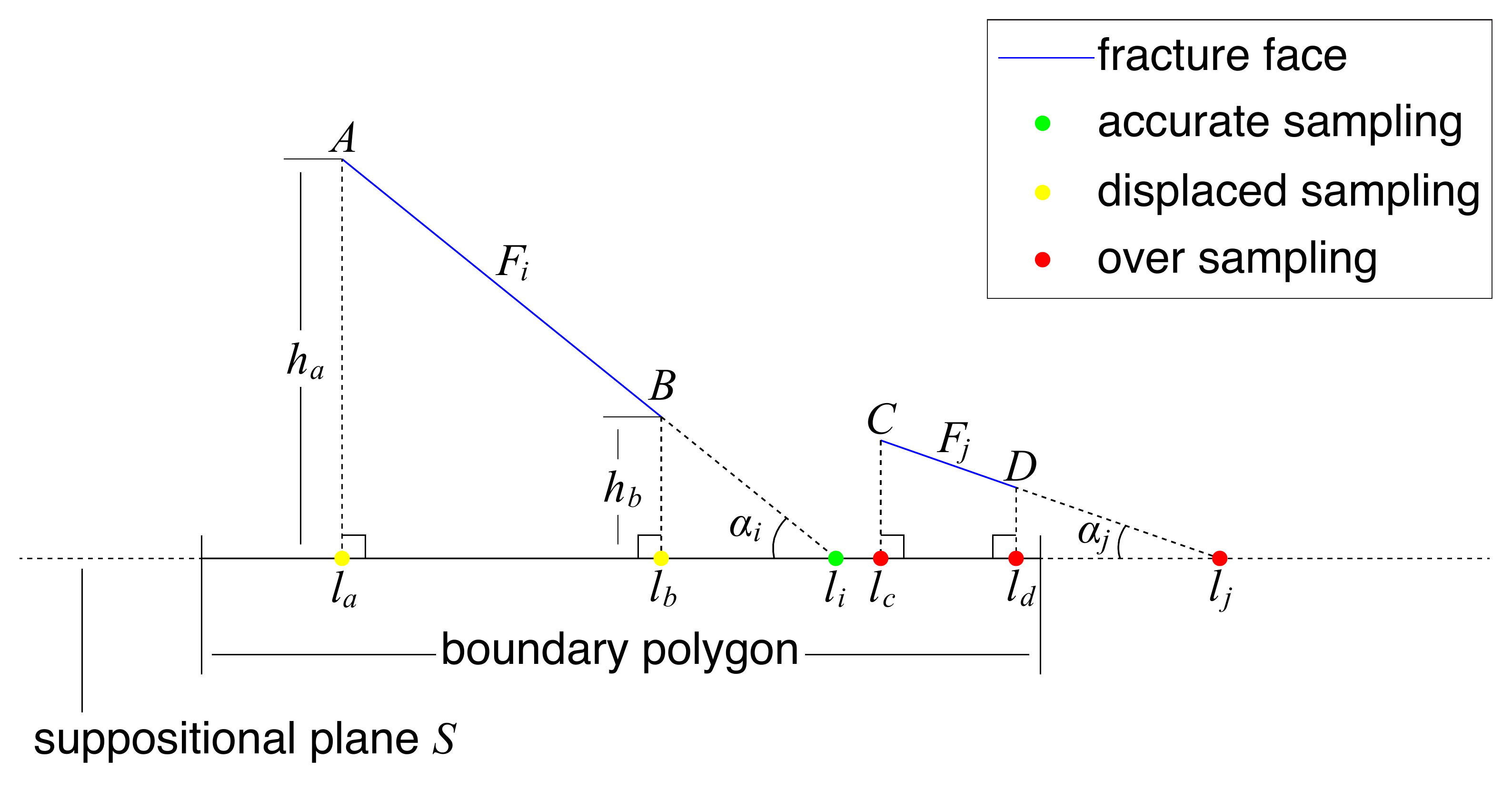}
  \caption{Illustration of estimating displacement error and over sampling error introduced by the conventional sampling methods. It's a side view of a 3D scene.}
  \label{fig:estimate_displacement_error}
\end{figure}

However, it's relatively easy to estimate the displacement error and the over sampling error introduced by the conventional sampling methods. Two situations need to be considered, as shown in \prettyref{fig:estimate_displacement_error}. It's a side view of a 3D scene, and the blue lines indicate fracture faces. Let $S$ be the suppositional plane. Fracture face $F_i$ intersects $S$ inside the boundary polygon with the intersection $l_i$. And $\alpha_i$ is the angle between $S$ and $F_i$. The conventional sampling methods will most likely sample at fracture edges, such as $A$ and $B$ in \prettyref{fig:estimate_displacement_error}. Let $h_a$ and $h_b$ be the distance from $A$ to $S$ and $B$ to $S$ respectively, if sampled at $A$, the displacement is $h_a/\tan{\alpha_i}$; if sampled at $B$, the displacement is $h_b/\tan{\alpha_i}$. The displacement of fracture face $F_i$ is $d_i = (h_a/\tan{\alpha_i} + h_b/\tan{\alpha_i})/2$ if $A$ and $B$ have the same probability to be sampled at. Let $L(x)$ be the function that gives the length of $x$, and $\{l_i | i \in \{1,...,n\}\}$ contains all fracture traces that aren't outside the boundary polygon, then the amount of displacement error is estimated as $E_\mathrm{dis} = \sum_{i=1}^n d_i / \sum_{i=1}^n L(l_i)$.

Fracture face $F_j$ intersects $S$ outside the boundary polygon with the intersection $l_j$, and $\{l_j | j \in \{1,...,k\}\}$ contains all fracture traces that are outside the boundary polygon. Then the amount of over sampling error is estimated as $E_\mathrm{os} = \sum_{j=1}^k L(l_j) / \sum_{i=1}^n L(l_i)$.

\begin{figure}[H]
  \centering
  \includegraphics[width=0.98\textwidth]{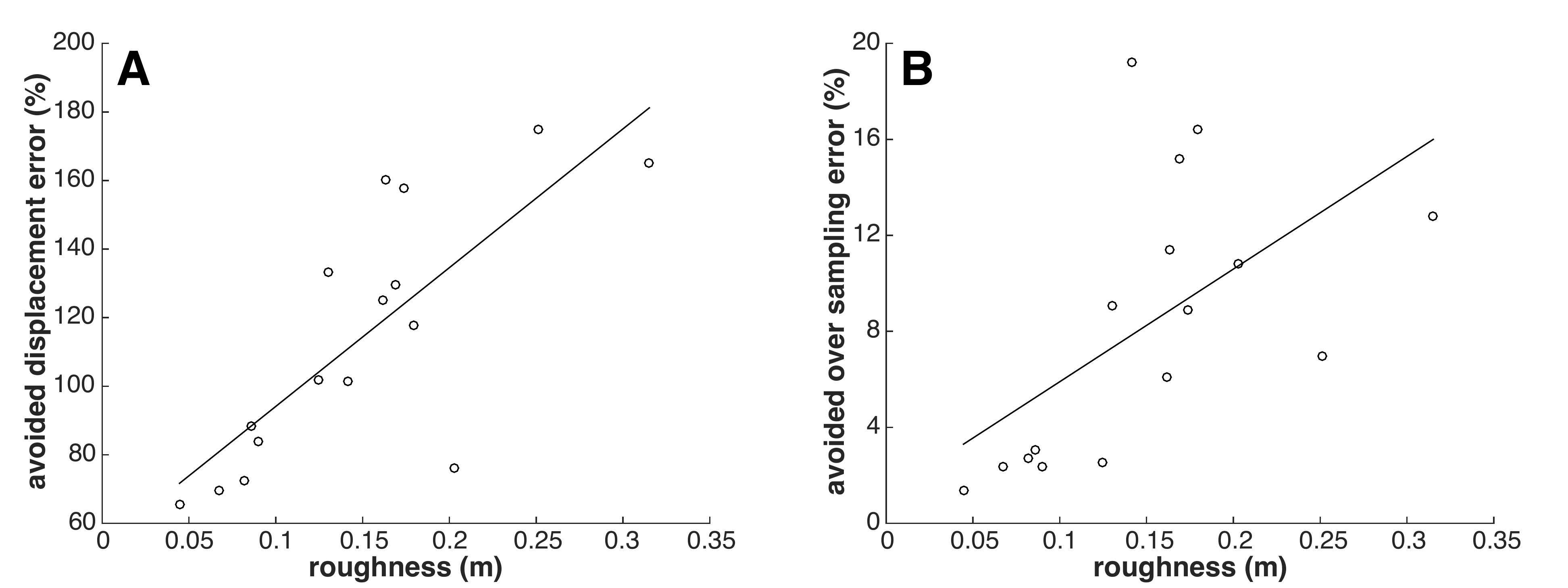}
  \caption{A: The displacement error (\%) avoided by the proposed method on outcrops of different roughnesses. B: The over sampling error (\%) avoided by the proposed method on outcrops of different roughnesses. Data collected from 16 outcrops.}
  \label{fig:error_analysis_plot}
\end{figure}

The amount of displacement error and over sampling error introduced by the conventional sampling methods are estimated on 16 outcrops with different roughnesses, and are plotted in \prettyref{fig:error_analysis_plot}. They are also sampling errors that can be avoided if our method was performed on those uneven outcrops, hence the advantages of the proposed method is demonstrated.

In \prettyref{fig:error_analysis_plot}, the roughness of the outcrop surface is defined as the mean distance from points on the surface ($\boldsymbol{p}_i \in P$) to its best fitting plane $S$. Both \prettyref{fig:error_analysis_plot}A and \prettyref{fig:error_analysis_plot}B show a tendency that sampling errors introduced by the conventional sampling methods grow with the increase of outcrop roughness. The displacement error (ranging from 60\% to 180\%) is more significant than the over sampling error (ranging from 1\% to 20\%). For outcrops other than bedding surfaces, or outcrops whose roughness $>$ 0.04 m, displacement error of at least 60\% can be introduced by the conventional sampling methods. For outcrops whose roughness $>$ 0.08 m, over sampling error may vary dramatically while growing with the increase of outcrop roughness.

\section{Conclusion}

The development of techniques that obtain DOMs from outcrops and extract the full extent of individual fractures offers the opportunity to address the problem of sampling errors introduced by performing the conventional sampling methods on uneven outcrop surfaces. In this paper, we have proposed a new method that performs outcrop fracture characterization on suppositional planes cutting through DOMs. The suppositional plane is the best fit plane of the outcrop surface, and the fracture trace map is extracted on the suppositional plane so that further fracture network characterization can be performed.

We compared the fracture trace map obtained from the conventional sampling methods and the sampling plane cutting thought the outcrop surface, and find that three main types of sampling errors introduced by the conventional sampling methods can be avoided by the proposed method: displacement error, orientation error and over sampling error. The amount of displacement error and over sampling error introduced by the conventional sampling methods on 16 outcrops with different roughnesses are estimated and analyzed. The results show that the conventional sampling methods don't apply to outcrops other than bedding surfaces or outcrops whose roughness $>$ 0.04 m, and demonstrates the advantages of our method of being able to avoid those sampling errors, greatly extend the types of outcrop surfaces for outcrop fracture characterization and increase the representativeness of the collected data to characterize the fracture network in outcrops.

\section*{Acknowledgement}

This study was funded by the Chinese National Science and Technology Major Project (2017ZX05008-001). We are grateful to Prof. Changjiang Li for valuable comments on earlier drafts.


\bibliography{\jobname.bib}

\begin{thebibliography}{17}
\expandafter\ifx\csname natexlab\endcsname\relax\def\natexlab#1{#1}\fi
\providecommand{\url}[1]{\texttt{#1}}
\providecommand{\href}[2]{#2}
\providecommand{\path}[1]{#1}
\providecommand{\DOIprefix}{doi:}
\providecommand{\ArXivprefix}{arXiv:}
\providecommand{\URLprefix}{URL: }
\providecommand{\Pubmedprefix}{pmid:}
\providecommand{\doi}[1]{\href{http://dx.doi.org/#1}{\path{#1}}}
\providecommand{\Pubmed}[1]{\href{pmid:#1}{\path{#1}}}
\providecommand{\bibinfo}[2]{#2}
\ifx\xfnm\relax \def\xfnm[#1]{\unskip,\space#1}\fi
\bibitem[{Adler and Thovert(1999)}]{Adler99}
\bibinfo{author}{Adler, P.M.}, \bibinfo{author}{Thovert, J.},
  \bibinfo{year}{1999}.
\newblock \bibinfo{title}{Fractures and fracture networks}.
  volume~\bibinfo{volume}{15}.
\newblock \bibinfo{publisher}{Springer Science \& Business Media}.
\bibitem[{Barber et~al.(1996)Barber, Dobkin and Huhdanpaa}]{Barber96}
\bibinfo{author}{Barber, C.B.}, \bibinfo{author}{Dobkin, D.P.},
  \bibinfo{author}{Huhdanpaa, H.}, \bibinfo{year}{1996}.
\newblock \bibinfo{title}{The quickhull algorithm for convex hulls}.
\newblock \bibinfo{journal}{ACM Transactions on Mathematical Software (TOMS)}
  \bibinfo{volume}{22}, \bibinfo{pages}{469--483}.
\bibitem[{Bellian et~al.(2005)Bellian, Kerans and Jennette}]{Bellian05}
\bibinfo{author}{Bellian, J.A.}, \bibinfo{author}{Kerans, C.},
  \bibinfo{author}{Jennette, D.C.}, \bibinfo{year}{2005}.
\newblock \bibinfo{title}{{Digital Outcrop Models: Applications of Terrestrial
  Scanning Lidar Technology in Stratigraphic Modeling}}.
\newblock \bibinfo{journal}{Journal of Sedimentary Research}
  \bibinfo{volume}{75}, \bibinfo{pages}{166--176}.
\bibitem[{Berkmann and Caelli(1994)}]{Berkmann94}
\bibinfo{author}{Berkmann, J.}, \bibinfo{author}{Caelli, T.},
  \bibinfo{year}{1994}.
\newblock \bibinfo{title}{{Computation of surface geometry and segmentation
  using covariance techniques}}.
\newblock \bibinfo{journal}{IEEE Transactions on Pattern Analysis and Machine
  Intelligence} \bibinfo{volume}{16}, \bibinfo{pages}{1114--1116}.
\bibitem[{Mah et~al.(2013)Mah, Samson, McKinnon and Thibodeau}]{Mah13}
\bibinfo{author}{Mah, J.}, \bibinfo{author}{Samson, C.},
  \bibinfo{author}{McKinnon, S.D.}, \bibinfo{author}{Thibodeau, D.},
  \bibinfo{year}{2013}.
\newblock \bibinfo{title}{{3D laser imaging for surface roughness analysis}}.
\newblock \bibinfo{journal}{International Journal of Rock Mechanics and Mining
  Sciences} \bibinfo{volume}{58}, \bibinfo{pages}{111--117}.
\bibitem[{Mauldon et~al.(2001)Mauldon, Dunne and Rohrbaugh}]{Mauldon01}
\bibinfo{author}{Mauldon, M.}, \bibinfo{author}{Dunne, W.},
  \bibinfo{author}{Rohrbaugh, M.}, \bibinfo{year}{2001}.
\newblock \bibinfo{title}{Circular scanlines and circular windows: new tools
  for characterizing the geometry of fracture traces}.
\newblock \bibinfo{journal}{Journal of Structural Geology}
  \bibinfo{volume}{23}, \bibinfo{pages}{247--258}.
\bibitem[{McCaffrey et~al.(2005)McCaffrey, Jones, Holdsworth, Wilson, Clegg,
  Imber, Holliman and Trinks}]{Mccaffrey05}
\bibinfo{author}{McCaffrey, K.}, \bibinfo{author}{Jones, R.},
  \bibinfo{author}{Holdsworth, R.}, \bibinfo{author}{Wilson, R.},
  \bibinfo{author}{Clegg, P.}, \bibinfo{author}{Imber, J.},
  \bibinfo{author}{Holliman, N.}, \bibinfo{author}{Trinks, I.},
  \bibinfo{year}{2005}.
\newblock \bibinfo{title}{{Unlocking the spatial dimension: digital
  technologies and the future of geoscience fieldwork}}.
\newblock \bibinfo{journal}{Journal of the Geological Society}
  \bibinfo{volume}{162}, \bibinfo{pages}{927--938}.
\bibitem[{Olariu et~al.(2008)Olariu, Ferguson, Aiken and Xu}]{Olariu08}
\bibinfo{author}{Olariu, M.I.}, \bibinfo{author}{Ferguson, J.F.},
  \bibinfo{author}{Aiken, C.L.}, \bibinfo{author}{Xu, X.},
  \bibinfo{year}{2008}.
\newblock \bibinfo{title}{{Outcrop fracture characterization using terrestrial
  laser scanners: Deep-water Jackfork sandstone at Big Rock Quarry, Arkansas}}.
\newblock \bibinfo{journal}{Geosphere} \bibinfo{volume}{4},
  \bibinfo{pages}{247--259}.
\bibitem[{Pahl(1981)}]{Pahl81}
\bibinfo{author}{Pahl, P.}, \bibinfo{year}{1981}.
\newblock \bibinfo{title}{Estimating the mean length of discontinuity traces},
  in: \bibinfo{booktitle}{International Journal of Rock Mechanics and Mining
  Sciences \& Geomechanics Abstracts}, \bibinfo{organization}{Elsevier}. pp.
  \bibinfo{pages}{221--228}.
\bibitem[{Pollyea and Fairley(2011)}]{Pollyea11}
\bibinfo{author}{Pollyea, R.M.}, \bibinfo{author}{Fairley, J.P.},
  \bibinfo{year}{2011}.
\newblock \bibinfo{title}{Estimating surface roughness of terrestrial laser
  scan data using orthogonal distance regression}.
\newblock \bibinfo{journal}{Geology} \bibinfo{volume}{39},
  \bibinfo{pages}{623--626}.
\bibitem[{Priest and Hudson(1981)}]{Priest81}
\bibinfo{author}{Priest, S.}, \bibinfo{author}{Hudson, J.},
  \bibinfo{year}{1981}.
\newblock \bibinfo{title}{Estimation of discontinuity spacing and trace length
  using scanline surveys}, in: \bibinfo{booktitle}{International Journal of
  Rock Mechanics and Mining Sciences \& Geomechanics Abstracts},
  \bibinfo{organization}{Elsevier}. pp. \bibinfo{pages}{183--197}.
\bibitem[{Priest(1993)}]{Priest93}
\bibinfo{author}{Priest, S.D.}, \bibinfo{year}{1993}.
\newblock \bibinfo{title}{Discontinuity Analysis for Rock Engineering}.
\newblock \bibinfo{publisher}{Springer Science \& Business Media}.
\bibitem[{Rohrbaugh~Jr et~al.(2002)Rohrbaugh~Jr, Dunne and
  Mauldon}]{Rohrbaugh02}
\bibinfo{author}{Rohrbaugh~Jr, M.}, \bibinfo{author}{Dunne, W.},
  \bibinfo{author}{Mauldon, M.}, \bibinfo{year}{2002}.
\newblock \bibinfo{title}{Estimating fracture trace intensity, density, and
  mean length using circular scan lines and windows}.
\newblock \bibinfo{journal}{AAPG Bulletin} \bibinfo{volume}{86},
  \bibinfo{pages}{2089--2104}.
\bibitem[{Sanderson and Nixon(2015)}]{Sanderson15}
\bibinfo{author}{Sanderson, D.J.}, \bibinfo{author}{Nixon, C.W.},
  \bibinfo{year}{2015}.
\newblock \bibinfo{title}{The use of topology in fracture network
  characterization}.
\newblock \bibinfo{journal}{Journal of Structural Geology}
  \bibinfo{volume}{72}, \bibinfo{pages}{55--66}.
\bibitem[{Wang et~al.(2017)Wang, Zou, Shen, Ren and Qin}]{Wang17}
\bibinfo{author}{Wang, X.}, \bibinfo{author}{Zou, L.}, \bibinfo{author}{Shen,
  X.}, \bibinfo{author}{Ren, Y.}, \bibinfo{author}{Qin, Y.},
  \bibinfo{year}{2017}.
\newblock \bibinfo{title}{A region-growing approach for automatic outcrop
  fracture extraction from a three-dimensional point cloud}.
\newblock \bibinfo{journal}{Computers \& Geosciences} \bibinfo{volume}{99},
  \bibinfo{pages}{100--106}.
\newblock \DOIprefix\doi{10.1016/j.cageo.2016.11.002}.
\bibitem[{Wu and Pollard(1995)}]{Wu95}
\bibinfo{author}{Wu, H.}, \bibinfo{author}{Pollard, D.D.},
  \bibinfo{year}{1995}.
\newblock \bibinfo{title}{An experimental study of the relationship between
  joint spacing and layer thickness}.
\newblock \bibinfo{journal}{Journal of Structural Geology}
  \bibinfo{volume}{17}, \bibinfo{pages}{887--905}.
\bibitem[{Xu et~al.(2000)Xu, Aiken, Bhattacharya, Corbeanu, Nielsen, McMechan
  and Abdelsalam}]{Xu00}
\bibinfo{author}{Xu, X.}, \bibinfo{author}{Aiken, C.L.V.},
  \bibinfo{author}{Bhattacharya, J.P.}, \bibinfo{author}{Corbeanu, R.M.},
  \bibinfo{author}{Nielsen, K.C.}, \bibinfo{author}{McMechan, G.A.},
  \bibinfo{author}{Abdelsalam, M.G.}, \bibinfo{year}{2000}.
\newblock \bibinfo{title}{{Creating virtual 3-D outcrop}}.
\newblock \bibinfo{journal}{The Leading Edge} \bibinfo{volume}{19},
  \bibinfo{pages}{197--202}.

\end{thebibliography}

\end{document}